\documentclass[%
 reprint,
 amsmath,amssymb,
 aps,
]{revtex4-2}

\usepackage{textgreek} 
\usepackage{upgreek}
\usepackage{lipsum}     
\usepackage[dvipsnames]{xcolor} 
\usepackage{gensymb}

\usepackage{graphicx}
\usepackage{dcolumn}
\usepackage{bm}

\begin{document}

\preprint{APS/123-QED}

\title{Gamma-Ray Flash Generation in Irradiating Thin Foil Target by Single Cycle Tightly Focused Extreme Power Laser Pulse}

\author{P. Hadjisolomou}
\email{Prokopis.Hadjisolomou@eli-beams.eu}
\author{T. M. Jeong}
\author{P. Valenta}
\altaffiliation{Faculty of Nuclear Sciences and Physical Engineering, Czech Technical University in Prague, Brehova 7, Prague 11519, Czech Republic}
\author{G. Korn}
\author{S. V. Bulanov}
\altaffiliation{National Institutes for Quantum and Radiological Science and Technology (QST), Kansai Photon Science Institute, 8-1-7 Umemidai, Kizugawa, Kyoto 619-0215, Japan}
\affiliation{
ELI Beamlines Centre, Institute of Physics, Czech Academy of Sciences, Za Radnicí 835, 25241 Dolní Břežany, Czech Republic
}

\date{\today}

\begin{abstract}
We present a regime where an ultra-intense laser pulse interacting with a foil target results in high \textgamma-photon conversion efficiency, obtained via three-dimensional quantum-electrodynamics particle-in-cell simulations. A single-cycle laser pulse is used under the tight-focusing condition for obtaining the $\mathrm{\lambda}^3$ regime. The simulations employ a radially polarized laser as it results in higher \textgamma-photon conversion efficiency compared to both azimuthal and linear polarizations. A significant fraction of the laser energy is transferred to positrons, while a part of the electromagnetic wave escapes the target as attosecond single-cycle pulses.
\end{abstract}

\maketitle



\par Since the invention of the Chirped Pulse Amplification \cite{1985_StricklandD} the laser peak power has increased to several petawatt (PW) \cite{2019_DansonC}. Recently, the demonstration of a $10 \kern0.2em \mathrm{PW}$ laser has been announced by the ELI-NP \cite{2020_TanakaKA}, while another one is near completion in the ELI-Beamlines. Furthermore, a $100 \kern0.2em \mathrm{PW}$ laser is currently under development \cite{2021_LiZ}. In parallel, independent efforts are made on the reduction of the pulse duration to the one wavelength level \cite{2006_BulanovSS, 2013_VoroninAA, 2020_OuilleM}.

\par The interaction of ultra-intense (${>} \kern0.1em 10^{22} \kern0.2em \mathrm{W cm^{-2}}$) laser pulses with matter results in the generation of high energy photons, protons/ions and energetic electrons/positrons \cite{2006_MourouG}. The generation and implementation of high energy \textgamma-photons is one of the main goals of multi-PW laser facilities. Notably, they can find direct application in fundamental studies such as in astrophysics \cite{2015_BulanovSV, 1992_ReesMJ, 2018_PhilippovAA}, materials science at extreme energy density \cite{2013_EliassonB}, and photonuclear reactions \cite{2004_NedorezovVG}.

\par At intensities above ${\sim} \kern0.1em 10^{24} \kern0.2em \mathrm{W cm^{-2}}$ the multiphoton Compton scattering radiation \cite{2013_RidgersCP, 2018_LezhninKV} dominates; in the present work we consider intensities of $10^{25} \kern0.2em \mathrm{W cm^{-2}}$. Photon emission through the multiphoton Compton scattering requires the collision of the laser pulse with an electron ($e^{-}$) or positron ($e^{+}$), resulting in a scattered \textgamma-photon through the process, $\mathrm{e^{\pm}} + N \omega_l \rightarrow \mathrm{e^{\pm}} + \omega_\gamma$, where $N >> 1$ is the photon number, $\omega_l$ is the central laser frequency, and $\omega_\gamma$ is the scattered \textgamma-photon frequency.

\begin{figure}
  \centering
  \includegraphics[width=1.0\linewidth]{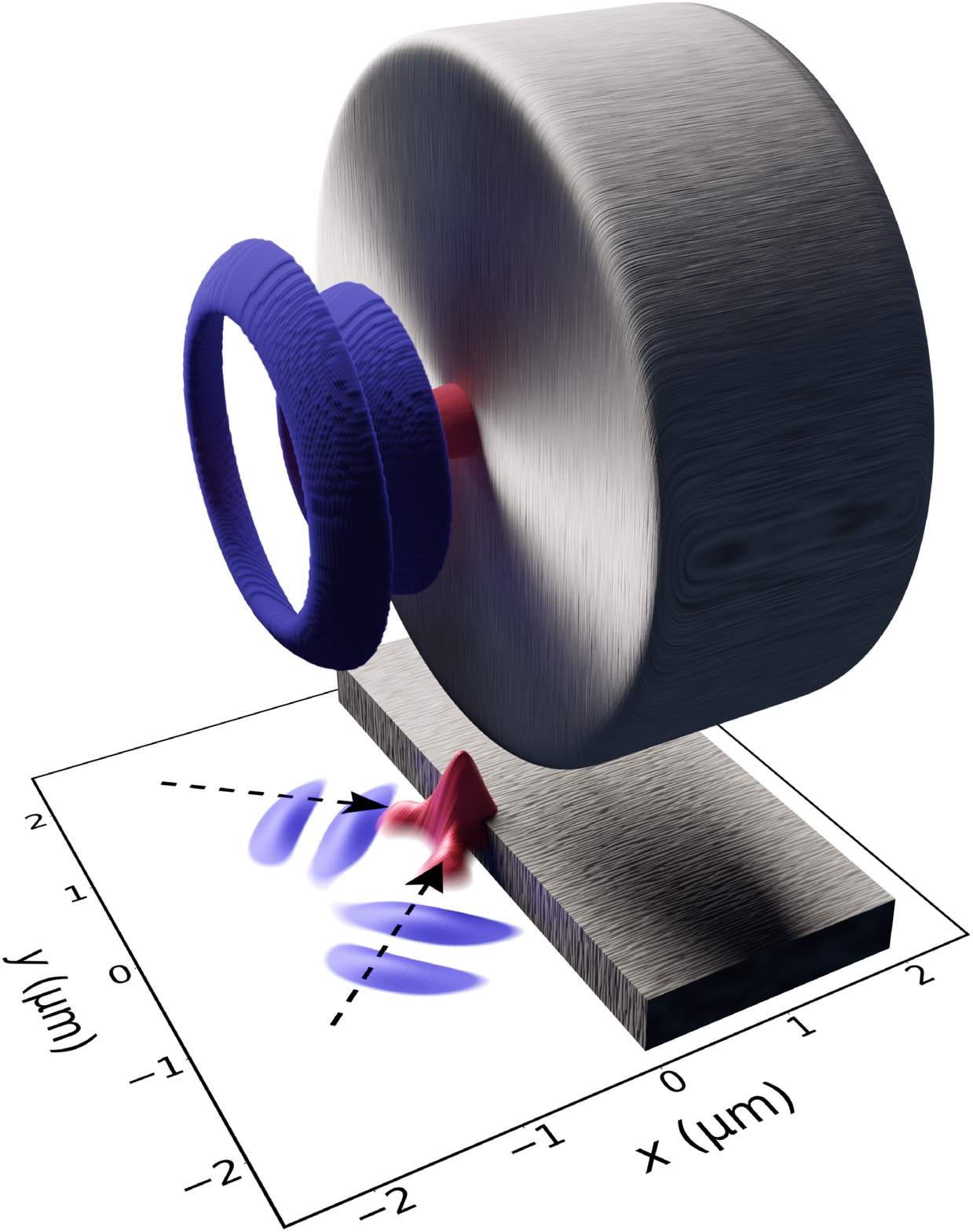}
  \caption{A 3D representation of the simulation setup. The gray cylinder shows the target. The blue isosurface shows the laser pulse at $2 \kern0.1em {\times} \kern0.1em 10^{24} \kern0.2em \mathrm{W cm^{-2}}$, as imported in EPOCH, at ${\sim} \kern0.1em - 4.3 \kern0.2em \mathrm{fs}$. The red isosurface, at $10^{25} \kern0.2em \mathrm{W cm^{-2}}$, is the time-averaged focused laser pulse. The contour plot shows the $(x,y)$-plane of the 3D configuration at $z=0$.}
  \label{fig:1}
\end{figure}

\par The \textgamma-photon emission is governed by the parameter $\chi_e =  \gamma_e \sqrt{ ({\bf{E}} + {\bf{v}} \times {\bf{B}})^2 - ({\bf{v}} \cdot {\bf{E}} / c)^2 } $ \cite{1970_RitusVI}, where $\gamma_e$ is the Lorentz factor of the electron/positron with velocity $\bf{v}$ before scattering, c is the speed of light in vacuum, ${\bf{E}}$ is the electric field normalized to $E_S$ ($E_S = m_e^2 c^3 / (e \hbar) \approx 1.3 \times 10^{18} \kern0.2em \mathrm{V m^{-1}}$ is the Schwinger field, $m_e$ is the electron mass, $\hbar$ is the reduced Planck constant, $e$ is the elementary charge), and ${\bf{B}}$ is the magnetic field normalized to $E_S$. When $\chi_e >> 1$ then a significant laser energy fraction is transferred to the \textgamma-photons \cite{2012_NakamuraT, 2012_RidgersCP}.

\par At the intensity levels presented below, electron-positron pairs are created via the multiphoton Breit-Wheeler process \cite{2009_EhlotzkyF}, where a high energy photon interacts with $N$ photons, ${ \omega_\gamma + N \omega_l \rightarrow \mathrm{e^{-}} + \mathrm{e^{+}} }$. This process is governed by the parameter $\chi_\gamma =  (\hbar \omega_l /  m_e c^2) \sqrt{ ({\bf{E}} + c {\bf{\hat{k}}} \times {\bf{B}})^2 - ({\bf{\hat{k}} \cdot \bf{E}})^2 } $, where $\bf{\hat{k}}$ is the unit vector of the \textgamma-photon propagation direction \cite{1970_RitusVI}. Several schemes have been proposed to optimize the laser to \textgamma-photon energy conversion efficiency \cite{2016_VranicM, 2017_GongZ, 2019_MagnussonJ, 2021_SampathA}.

\par The focused intensity of $10^{25} \kern0.2em \mathrm{W cm^{-2}}$ corresponds to a peak electric field of ${\approx} \kern0.1em 8.7 \kern0.1em {\times} \kern0.1em 10^{15} \kern0.2em \mathrm{V m^{-1}}$. Assuming the laser as a $1 \kern0.2em \mathrm{\upmu m}$ wavelength plane wave, the dimensionless amplitude is $a_0 = e E / (m_e c \omega_l) \approx 2700 >> 1$ giving an electron Lorentz factor of $\gamma_e \approx a_0$.


\par In this letter we introduce a regime where the laser to \textgamma-photons conversion efficiency approaches $50 \kern0.2em \%$, resulting in a ${\sim} \kern0.1em 40 \kern0.2em \mathrm{PW}$ \textgamma-ray flash generation. This regime employs the tight-focusing scheme \cite{2015_JeongTM, 2018_JeongTM} with a radially polarized (transverse magnetic mode) laser and near-single-cycle pulse corresponding to the $\mathrm{\lambda}^3$ regime \cite{2002_MourouG} (where $\lambda$ is the laser wavelength), and a time-average intensity of $10^{25} \kern0.2em \mathrm{W cm^{-2}}$ is reached by an ${\sim} \kern0.1em 80 \kern0.2em \mathrm{PW}$ laser. We further consider the cases for linearly and azimuthally polarized (transverse electric) lasers, demonstrating a strong enhancement of the conversion efficiency through radial polarization. In all cases a significant energy fraction is transferred to electron-positron pairs. A portion of the electromagnetic (EM) field energy escapes the target as distinct attosecond pulses (see also Ref. \cite{2004_NaumovaNM}).


\par In order to obtain the three-dimensional (3D) arrays containing the electric and magnetic focusing fields under the tight-focusing condition, we developed a dedicated Fortran code, following the theoretical description in Ref. \cite{2015_JeongTM}. The code is adjusted for both radially and azimuthally polarized lasers by replacing the field integrands in Ref. \cite{2015_JeongTM} with those described in Ref. \cite{2018_JeongTM}. The field data arrays are imported into a particle-in-cell (PIC) code, propagating the unfocused fields, which are focused at a reference time of $t=0  \kern0.2em \mathrm{fs}$. The $\mathrm{\lambda}^3$ regime requires a strong localization of the laser pulse in narrow temporal and spatial domains. The temporal profile is simulated by superposing a series of EM waves with wavelengths within the range $(700-1750) \kern0.2em \mathrm{nm}$, resulting in a near-single-cycle pulse with a full-width-at-half-maximum of the pulse duration envelope of ${\sim} \kern0.1em 3.6 \kern0.2em \mathrm{fs}$ (it is slightly larger than the central wavelength of $1 \kern0.2em \mathrm{\upmu m}$). The f-number (focal length divided by the beam diameter) used is $1/3$. The f-number ensures the tight-focusing for achieving the $\mathrm{\lambda}^3$ regime. The coupling of the externally imported fields with a PIC code is realized with the setup shown in Fig. \ref{fig:1}.

\begin{figure}
  \centering
  \includegraphics[width=1.0\linewidth]{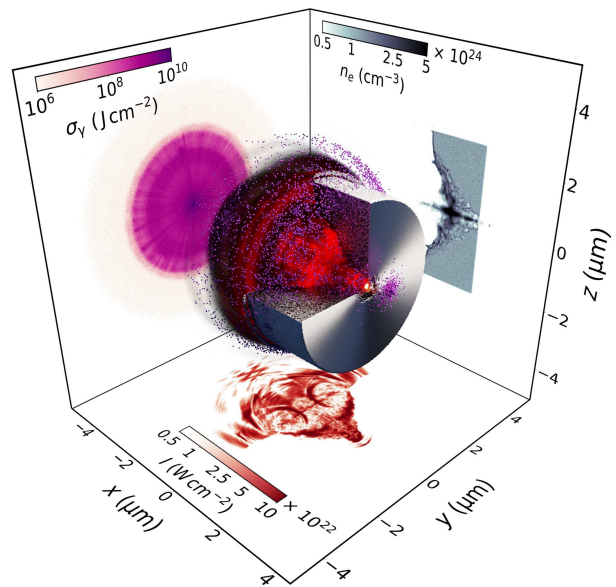}
  \caption{A 3D representation of the simulation results at ${\sim} \kern0.1em 6.7 \kern0.2em \mathrm{fs}$. The intensity and electron density are shown by red and gray colors, respectively; the purple dots are the \textgamma-photon macroparticles with energy $\ge 0.75 \kern0.2em \mathrm{GeV}$. The $(x,y)$-plane and $(x,z)$-plane at $z=0$ show the cross-sections of the intensity and electron density. The \textgamma-photon surface energy density is is shown on the $(y,z)$-plane.}
  \label{fig:2}
\end{figure}

\par The simulation is performed by the 3D EPOCH \cite{2015_ArberTD} PIC code. The code is compiled with the flags for the quantum electrodynamics \cite{2014_RidgersCP} and Higuera-Cary \cite{2017_HigueraAV} preprocessor directives enabled. The focusing laser fields are pre-calculated externally (at ${\sim} \kern0.1em - 4.3 \kern0.2em \mathrm{fs}$) and then imported into EPOCH. The simulation box edges of $10.24 \kern0.2em \mathrm{\upmu m}$ are split into 512 cells per dimension, resulting in cells of $\alpha_c = 20 \kern0.2em \mathrm{nm}$ edge. The simulation runs for  $16 \kern0.2em \mathrm{fs}$, i.e. before neither fields nor particles could exit the simulation box. The simulation box is large enough for the conversion of all species to saturate.

\par The target is a cylinder of height, $l$, with the front surface at $x=0$. The radius, $r=2.4 \kern0.2em \mathrm{\upmu m}$, is large enough for the target edges not to be destroyed  by the laser. The target consists of uniformly overlapping electrons and ions (i.e., no pre-plasma), with $8 \pi r^2 l / \alpha_c$ macroparticles (8 per cell) assigned to each species. The ion atomic number is set to unity and the mass number to 2.2, where 2.2 is the average mass to charge ratio for solid elements with atomic number less than 50. This approach in the EPOCH gives identical results as if using unscaled atomic and mass numbers, enabling generalization of the results for a broad collection of target materials. In our parametric study, the target thickness varies within the range $0.2 \kern0.2em \mathrm{\upmu m} - 2 \kern0.2em \mathrm{\upmu m}$. The electron density is in the range $2 \times 10^{23} \kern0.2em \mathrm{cm^{-3}} - 5 \times 10^{24} \kern0.2em \mathrm{cm^{-3}}$. The relativistically modified skin depth is resolved with an accuracy of more than 10 cells at an average electron density of $10^{24} \kern0.2em \mathrm{cm^{-3}}$. Photons with energy less than $1 \kern0.2em \mathrm{MeV}$ are ignored because they account for ${\sim} \kern0.1em 1 \kern0.2em \%$ of the total photon energy.


\par An illustrative outcome of the simulations is presented in Fig. \ref{fig:2}, which shows the laser intensity along with the target electron density at ${\sim} \kern0.1em 6.7 \kern0.2em \mathrm{fs}$. The corresponding target has a thickness of $2 \kern0.2em \mathrm{\upmu m}$ and electron density of $1.2 \times 10^{24} \kern0.2em \mathrm{cm^{-3}}$. As can be seen in Fig. \ref{fig:2}, a conical channel is drilled by the laser pulse in the target. Furthermore, a low electron density is formed on the target front region.

\begin{figure}
  \centering
  \includegraphics[width=1.0\linewidth]{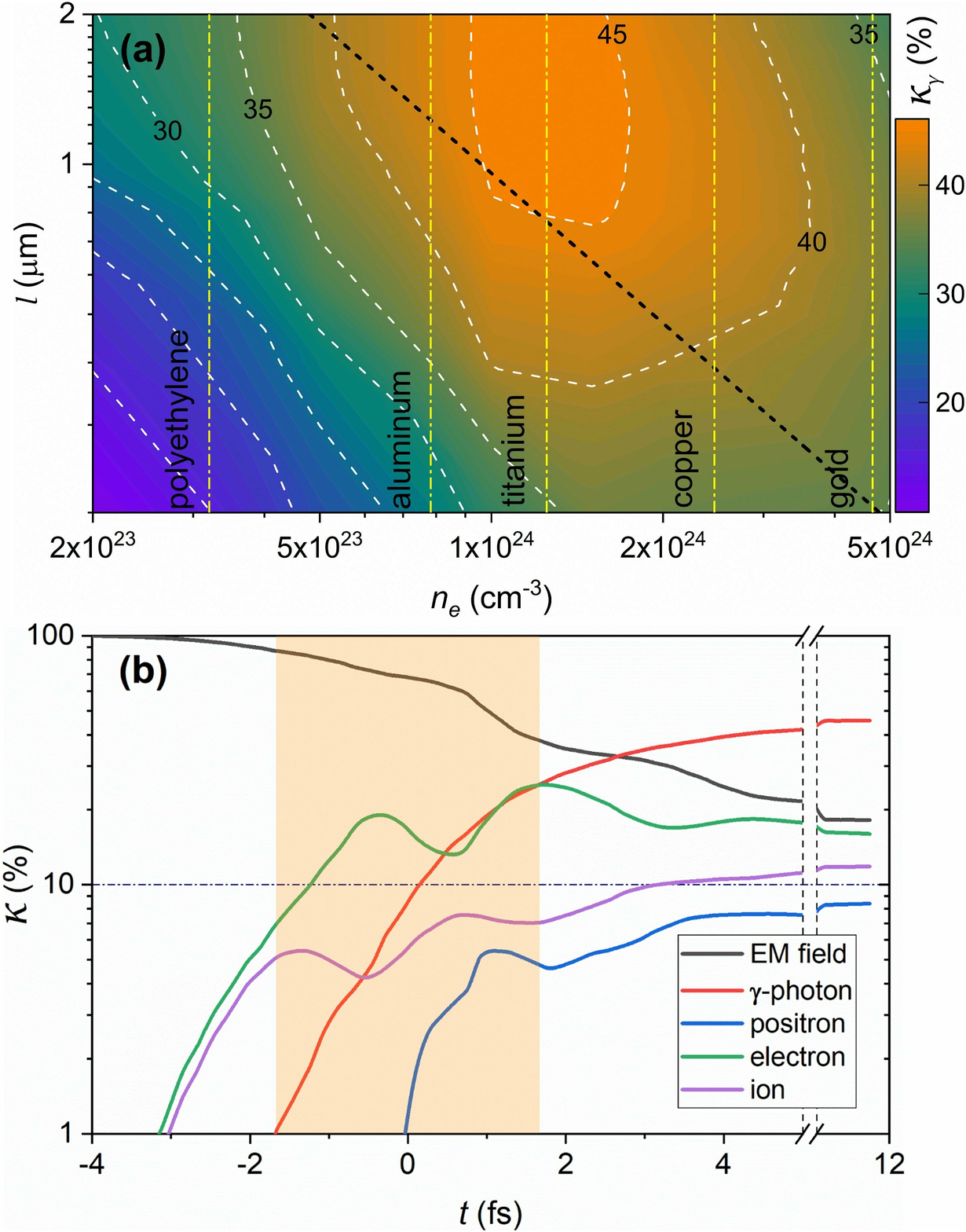}
  \caption{ (a) The laser to \textgamma-photon conversion efficiency as a function of the target thickness and electron density. (b) The laser to \textgamma-photon conversion efficiency versus time. The black, red, blue, green, and purple lines correspond to the EM-field, \textgamma-photons, positrons, electrons, and ions, respectively. The orange region corresponds to the laser period at the focus.}
  \label{fig:3}
\end{figure}

\par The key factor for the formation of the conical channel structure is the extremely tightly-focused radially polarized laser. At the focus, especially on axis, the laser beam is characterized by a dominant longitudinal electric field component, $E_x$, compared to the transverse one, $E_r$ \cite{2018_JeongTM}, while the magnetic field is purely azimuthal, $B_\theta$. In contrary, in an azimuthally polarized laser the longitudinal electric field component vanishes. Therefore, one can consider $E_x$ as a driving field directly ejecting electrons from a region with the diameter of ${\sim} \kern0.1em \lambda/2$ and creating a channel through which the laser further penetrates the target. As a result, extremely high EM wave intensity is observed locally within the channel, even as high as ${\sim} \kern0.2em 10^{26} \kern0.2em \mathrm{W cm^{-2}}$; although strong energy conversion, mainly to the \textgamma-photons, results in a macroscopic intensity decrease.

\par The conversion of laser to the \textgamma-photon energy is illustrated in Fig. \ref{fig:3}. The multiphoton Compton scattering determining this process depends on the parameter $\chi_e$. In the case of tight-focusing the electric and magnetic fields are separated in space (${\bf{E_r}} \ne c {\bf{B_\theta}}$). As a result, at most locations, the electric and magnetic contributions to $\chi_e$ do not cancel, even for electrons or positrons moving along the laser propagation direction. Therefore, \textgamma-photon production is allowed in most regions where relativistic electrons coexist with either strong electric or magnetic field.

\par
\begin{figure}
  \centering
  \includegraphics[width=1.0\linewidth]{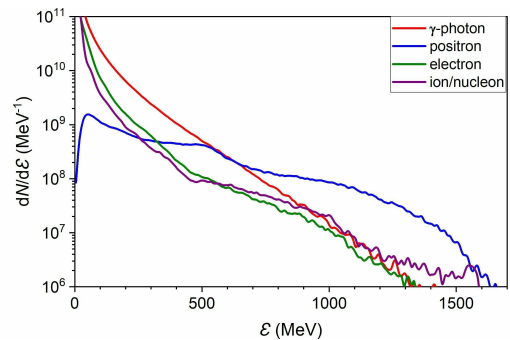}
  \caption{The energy spectra of \textgamma-photons, positrons, electrons, and ions (per nucleon) are shown with the red, blue, green, and purple colors, respectively. The simulation is for a $2 \kern0.2em \mathrm{\upmu m}$ thick foil with an electron density of $1.2 \kern0.1em {\times} \kern0.1em 10^{24} \kern0.2em \mathrm{cm^{-3}}$.}
  \label{fig:4}
\end{figure}

\begin{figure}
  \centering
  \includegraphics[width=1.0\linewidth]{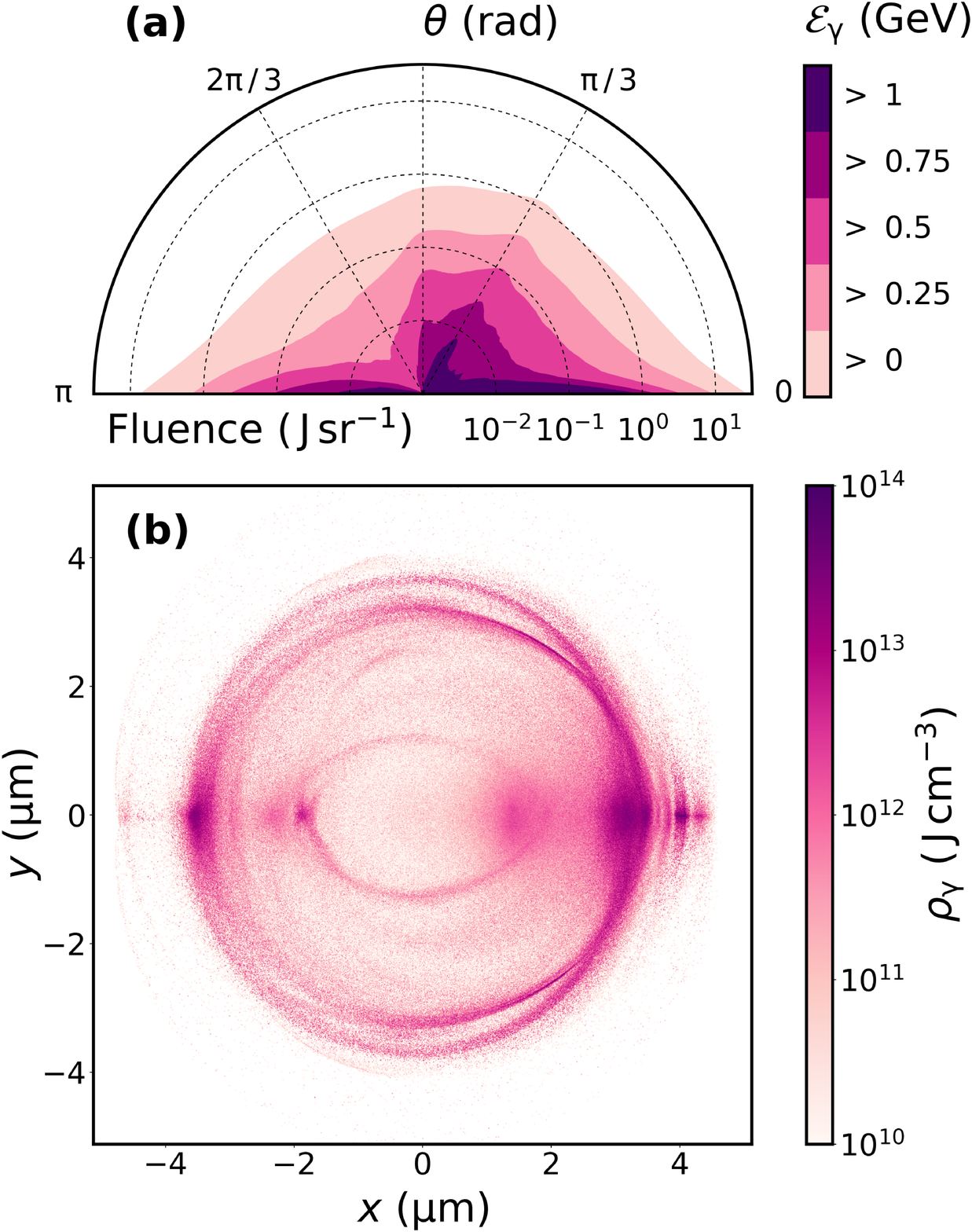}
  \caption{ (a) The angular distribution of the \textgamma-photon fluence. (b) The \textgamma-photon energy density in the $(x,y)$-plane at $z=0$.}
  \label{fig:5}
\end{figure}

 Our parametric study reveals a strong dependence of the laser to \textgamma-photon conversion efficiency on both the electron density and target thickness, as shown in Fig. \ref{fig:3}(a). The figure shows the isocontours of the laser to \textgamma-photon conversion efficiency, as a function of the target thickness and electron density. It is seen that the conversion efficiency increases with the target thickness, where the increase is steeper for higher electron densities. Furthermore, a sharp conversion efficiency increase is observed by increasing the electron density up to ${\sim} \kern0.1em 1.2 \times 10^{24} \kern0.2em \mathrm{cm^{-3}}$, followed by a gradual slow decrease, but not falling below $35 \kern0.2em \%$ for the higher densities simulated. At the optimum thickness-density configuration, the conversion efficiency to \textgamma-photons reaches ${\sim} \kern0.1em 47 \kern0.2em \%$, to positrons ${\sim} \kern0.1em 8 \kern0.2em \%$, to electrons ${\sim} \kern0.1em 16 \kern0.2em \%$, to ions ${\sim} \kern0.1em 12 \kern0.2em \%$ and only a ${\sim} \kern0.1em 17 \kern0.2em \%$ of the initial pulse remains in the form of the EM-field. The \textgamma-photon conversion efficiency corresponds to a ${\sim} \kern0.1em 130 \kern0.2em \mathrm{J}$ \textgamma-ray flash generation, as seen in Fig. \ref{fig:5}(b). The \textgamma-photon spectrum (red line in Fig. \ref{fig:4}) is of a temperature of ${\sim} \kern0.1em 0.14 \kern0.2em \mathrm{GeV}$ with a maximum energy of ${\sim} \kern0.1em 1.3 \kern0.2em \mathrm{GeV}$. It is estimated that each electron has emitted approximately $2.5$ \textgamma-photons by the end of the laser-target interaction.

\par The optimal parameters of the target thickness, $l$, and electron density, $n_e$, correspond to the condition $a_0 = \epsilon_p$, where $\epsilon_p$ is the normalized areal density. It equals $\epsilon_p = n_e e^2 l / (2 \epsilon_0 m_e \omega_l c) = \pi n_e l / (n_{cr} \lambda)$ \cite{1998_VshivkovVA}, where $n_{cr}= \epsilon_0 m_e \omega^2 / e^2$ is the critical electron density, and $\epsilon_0$ is the vacuum permittivity. The relationship is shown by the dashed black line in Fig. \ref{fig:3}(a) and it corresponds to the conversion efficiency to all particle species, not only \textgamma-photons. Metallic foils commonly used in laser-matter interaction experiments (aluminum, copper, and gold) fall within our simulation parameters, as labeled by the yellow dashed lines in Fig. \ref{fig:3}(a). Here, an optimum conversion efficiency is predicted for titanium, with an electron density of ${\sim} \kern0.1em 1.25 \times 10^{24} \kern0.2em \mathrm{cm^{-3}}$. Frequently used polymer materials, such as polyethylene, correspond to significantly lower \textgamma-photon conversion efficiency.

\par Initially, the background field is several orders of magnitude weaker than the focused field, however, it is strong enough to heat surface electrons and perturb the initially smooth target surface. At consequent times this effect magnifies due to the rapid intensity increase forming a pre-plasma. The relativistically under-dense target allows penetration of the field in the target, where at a certain depth (sub-micron scale) a thin over-dense electron pile-up is formed.

\par At $t = 0 \kern0.2em \mathrm{fs}$, the pre-plasma is shaped into a torus-like density pattern, where the torus major radius coincides with the focused laser radius. Since prior that time the pulse sees a practically void region (no electrons or positrons to interact with), most of the laser to \textgamma-photon energy conversion happens at the consequent times. For the simulation results shown in Fig. \ref{fig:3}(b), when the field reaches the focus, only ${\sim} \kern0.1em 8 \kern0.2em \%$ of the laser energy has been converted to the \textgamma-photons.

\par The efficient laser-particle coupling occurs in the quarter-period time-scale, as seen from the oscillating patterns in Fig. \ref{fig:3}(b), for the charged particles. The figure shows the percentage of laser energy transferred to each particle specie versus time. The focused intensity is maximized within a time-window of a single laser-cycle, as shown in Fig. \ref{fig:3}(b), during which, the ${\sim} \kern0.1em 2/3$ of laser to the \textgamma-photon energy conversion happens.

\par At the initial times of the simulation, the directional electron jets occur at $0 \degree, \kern0.1em {\sim} \kern0.1em 60 \degree$, and $180 \degree$, where $0 \degree$ is the direction of a vector normal to the target rear. The electrons at ${\sim} \kern0.1em 60 \degree$ are connected to the focusing condition by the maximum half angle (${\sim} \kern0.1em 58 \degree$) of the light cone formed by the focusing optic with a f-number of $1/3$. The ${\sim} \kern0.1em 60 \degree$ electron beams are pushed towards the focal spot from the pre-plasma region, but then scattered by the strongly focused $\mathrm{\lambda}^3$ field and their distribution becomes more isotropic. The forward moving electron population is formed at early times with a maximum energy exceeding $2 \kern0.2em \mathrm{GeV}$, which after a time of $\lambda/ (4 c)$ is slowed down to ${\sim} \kern0.1em 1 \kern0.2em \mathrm{GeV}$. The backward moving electron beam, although formed at later times, also reaches ${\sim} \kern0.1em 1 \kern0.2em \mathrm{GeV}$ energy by the end of the simulation. The electron spectrum is shown by the green curve in Fig. \ref{fig:4}, exhibiting two temperatures, of ${\sim} \kern0.1em 0.11 \kern0.2em \mathrm{GeV}$ for electron energy below ${\sim} \kern0.1em 0.45 \kern0.2em \mathrm{GeV}$ and of ${\sim} \kern0.1em 0.23 \kern0.2em \mathrm{GeV}$ for electrons with energy up-to ${\sim} \kern0.1em 1.3 \kern0.2em \mathrm{GeV}$. A very similar spectrum is observed for ions. The motion of those distinct electron populations is in connection with the locations where several attosecond single-cycle EM pulses are generated, as seen in Fig. \ref{fig:2}. Similar patterns have been reported in Ref. \cite{2004_NaumovaNM}. By the end of the interaction, these pulses contain ${\sim} \kern0.1em 17 \kern0.2em \%$ of the initial laser energy. Although our simulations are optimized for high \textgamma-photon yield, reduced target thickness  significantly enhances the maximum electron, positron, and ion energy.

\par The direction of the emitted photons falls within a small angle (and practically overlaps) with the electron generation during the initial simulation stage. As revealed by our simulations, the most energetic photons are emitted at the initial steps of the interaction, with photons produced at each following time-step having gradually reduced maximum energy, in connection with the gradual reduction of the field peak intensity. Note also that the distribution of the \textgamma-photon beams emitted per time-step is not isotropic. However, the time-averaged distribution of the lower energy \textgamma-photons is isotropic. A directional \textgamma-photon motion is observed for the high energy (${\sim} \kern0.1em 500 \kern0.2em \mathrm{MeV}$) \textgamma-photons, at $0 \degree, \kern0.1em {\sim} \kern0.1em 60 \degree$, and $180 \degree$, shown in Fig. \ref{fig:5}(a).

\par For comparison, the energy transferred to all particles for linearly and azimuthally polarized lasers is reduced by ${\sim} \kern0.1em 9 \kern0.2em \%$ and ${\sim} \kern0.1em 37 \kern0.2em \%$ respectively, compared to the radially polarized laser. However, the decrease in the \textgamma-photon conversion efficiency is ${\sim} \kern0.1em 19 \kern0.2em \%$ and ${\sim} \kern0.1em 56 \kern0.2em \%$ respectively. The increase in the laser absorption efficiency and consequently in the conversion efficiency to \textgamma-photons for a radially polarized laser can be explained by the electric field configuration. For the f-number of $1/3$ the longitudinal electric field component of the radially polarized laser is larger than in the linear polarization case \cite{2018_JeongTM}. For the azimuthally polarized laser the longitudinal component of the electric field vanishes. Therefore, electrons can be more efficiently accelerated/decelerated along the longitudinal axis resulting in higher \textgamma-photon yield in the radially polarized case.

\par Our simulations also reveal a decrease in the maximum \textgamma-photon energy over time. This reduction is due to the multi-photon Breit-Wheeler process, leading to creation of the electron-positron pairs. Their spectrum is shown by the blue color in Fig. \ref{fig:4}, having a temperature of ${\sim} \kern0.1em 0.32 \kern0.2em \mathrm{GeV}$. However, we are at a regime where this effect is still minor; although ${\sim} \kern0.1em 8.6 \kern0.2em \%$ of the laser energy is transferred to positrons, the most of this fraction does not originate from \textgamma-photons. Further increase in the laser power creates a pair production avalanche \cite{2008_BellAR} significantly reducing the \textgamma-photon conversion efficiency.

\par In conclusion, we demonstrate via 3D PIC simulations that an ${\sim} \kern0.1em 80 \kern0.2em \mathrm{PW}$ laser is capable to produce a ${\sim} \kern0.1em 40 \kern0.2em \mathrm{PW}$ \textgamma-ray flash. In order for ${\sim} \kern0.1em 50 \kern0.2em \%$ of the laser energy to be converted to \textgamma-photons, a radially polarized laser is used in the $\mathrm{\lambda}^3$ regime, through tight-focusing of a near-single-cycle pulse. The key difference of a radially versus an azimuthally polarized laser under the tight-focusing scheme is the presence of an extremely strong longitudinal electric field component for the former, which enhances the \textgamma-photon generation. Approximately $8 \kern0.2em \%$ of the laser energy is transferred to positrons, mainly under both direct acceleration by the laser and by the energy transfer from the escaping electrons. A fraction of ${\sim} \kern0.1em 17 \kern0.2em \%$ of the laser energy forms several attosecond pulses due to re-emission by energetic electrons formed during the initial stage of the laser-foil interaction.


\par The authors would like to acknowledge discussions with Dr. C. Ridgers regarding the quantum electrodynamics EPOCH module and useful communication with Dr. D. Khikhlukha and K. Lezhnin. This work is supported by the projects High Field Initiative (CZ.02.1.01/0.0/0.0/15\_003/0000449) from the European Regional Development Fund and IT4Innovations National Supercomputing Center (Grant No. LM2015070) from the Ministry of Education, Youth and Sports of the Czech Republic. The EPOCH code is in part funded by the UK EPSRC grants EP/G054950/1, EP/G056803/1, EP/G055165/1 and EP/M022463/1.

\bibliography{PRL_BIB} 

\providecommand{\noopsort}[1]{}\providecommand{\singleletter}[1]{#1}%
\begin{thebibliography}{32}%
\makeatletter
\providecommand \@ifxundefined [1]{%
 \@ifx{#1\undefined}
}%
\providecommand \@ifnum [1]{%
 \ifnum #1\expandafter \@firstoftwo
 \else \expandafter \@secondoftwo
 \fi
}%
\providecommand \@ifx [1]{%
 \ifx #1\expandafter \@firstoftwo
 \else \expandafter \@secondoftwo
 \fi
}%
\providecommand \natexlab [1]{#1}%
\providecommand \enquote  [1]{``#1''}%
\providecommand \bibnamefont  [1]{#1}%
\providecommand \bibfnamefont [1]{#1}%
\providecommand \citenamefont [1]{#1}%
\providecommand \href@noop [0]{\@secondoftwo}%
\providecommand \href [0]{\begingroup \@sanitize@url \@href}%
\providecommand \@href[1]{\@@startlink{#1}\@@href}%
\providecommand \@@href[1]{\endgroup#1\@@endlink}%
\providecommand \@sanitize@url [0]{\catcode `\\12\catcode `\$12\catcode
  `\&12\catcode `\#12\catcode `\^12\catcode `\_12\catcode `\%12\relax}%
\providecommand \@@startlink[1]{}%
\providecommand \@@endlink[0]{}%
\providecommand \url  [0]{\begingroup\@sanitize@url \@url }%
\providecommand \@url [1]{\endgroup\@href {#1}{\urlprefix }}%
\providecommand \urlprefix  [0]{URL }%
\providecommand \Eprint [0]{\href }%
\providecommand \doibase [0]{https://doi.org/}%
\providecommand \selectlanguage [0]{\@gobble}%
\providecommand \bibinfo  [0]{\@secondoftwo}%
\providecommand \bibfield  [0]{\@secondoftwo}%
\providecommand \translation [1]{[#1]}%
\providecommand \BibitemOpen [0]{}%
\providecommand \bibitemStop [0]{}%
\providecommand \bibitemNoStop [0]{.\EOS\space}%
\providecommand \EOS [0]{\spacefactor3000\relax}%
\providecommand \BibitemShut  [1]{\csname bibitem#1\endcsname}%
\let\auto@bib@innerbib\@empty
\bibitem [{\citenamefont {Strickland}\ and\ \citenamefont
  {Mourou}(1985)}]{1985_StricklandD}%
  \BibitemOpen
  \bibfield  {author} {\bibinfo {author} {\bibfnamefont {D.}~\bibnamefont
  {Strickland}}\ and\ \bibinfo {author} {\bibfnamefont {G.}~\bibnamefont
  {Mourou}},\ }\bibfield  {title} {\bibinfo {title} {Compression of amplified
  chirped optical pulses},\ }\href
  {https://doi.org/https://doi.org/10.1016/0030-4018(85)90120-8} {\bibfield
  {journal} {\bibinfo  {journal} {Opt. Commun.}\ }\textbf {\bibinfo {volume}
  {56}},\ \bibinfo {pages} {219} (\bibinfo {year} {1985})}\BibitemShut
  {NoStop}%
\bibitem [{\citenamefont {Danson}\ \emph {et~al.}(2019)\citenamefont {Danson},
  \citenamefont {Haefner}, \citenamefont {Bromage}, \citenamefont {Butcher},
  \citenamefont {Chanteloup}, \citenamefont {Chowdhury}, \citenamefont
  {Galvanauskas}, \citenamefont {Gizzi}, \citenamefont {Hein}, \citenamefont
  {Hillier},\ and\ \citenamefont {et~al.}}]{2019_DansonC}%
  \BibitemOpen
  \bibfield  {author} {\bibinfo {author} {\bibfnamefont {C.~N.}\ \bibnamefont
  {Danson}}, \bibinfo {author} {\bibfnamefont {C.}~\bibnamefont {Haefner}},
  \bibinfo {author} {\bibfnamefont {J.}~\bibnamefont {Bromage}}, \bibinfo
  {author} {\bibfnamefont {T.}~\bibnamefont {Butcher}}, \bibinfo {author}
  {\bibfnamefont {J.~F.}\ \bibnamefont {Chanteloup}}, \bibinfo {author}
  {\bibfnamefont {E.~A.}\ \bibnamefont {Chowdhury}}, \bibinfo {author}
  {\bibfnamefont {A.}~\bibnamefont {Galvanauskas}}, \bibinfo {author}
  {\bibfnamefont {L.~A.}\ \bibnamefont {Gizzi}}, \bibinfo {author}
  {\bibfnamefont {J.}~\bibnamefont {Hein}}, \bibinfo {author} {\bibfnamefont
  {D.~I.}\ \bibnamefont {Hillier}},\ and\ \bibinfo {author} {\bibnamefont
  {et~al.}},\ }\bibfield  {title} {\bibinfo {title} {Petawatt and exawatt class
  lasers worldwide},\ }\href {https://doi.org/10.1017/hpl.2019.36} {\bibfield
  {journal} {\bibinfo  {journal} {High Power Laser Sci.}\ }\textbf {\bibinfo
  {volume} {7}},\ \bibinfo {pages} {e54} (\bibinfo {year} {2019})}\BibitemShut
  {NoStop}%
\bibitem [{\citenamefont {Tanaka}\ \emph {et~al.}(2020)\citenamefont {Tanaka},
  \citenamefont {Spohr}, \citenamefont {Balabanski}, \citenamefont {Balascuta},
  \citenamefont {Capponi}, \citenamefont {Cernaianu}, \citenamefont {Cuciuc},
  \citenamefont {Cucoanes}, \citenamefont {Dancus}, \citenamefont {Dhal},
  \citenamefont {Diaconescu}, \citenamefont {Doria}, \citenamefont {Ghenuche},
  \citenamefont {Ghita}, \citenamefont {Kisyov}, \citenamefont {Nastasa},
  \citenamefont {Ong}, \citenamefont {Rotaru}, \citenamefont {Sangwan},
  \citenamefont {Söderström}, \citenamefont {Stutman}, \citenamefont
  {Suliman}, \citenamefont {Tesileanu}, \citenamefont {Tudor}, \citenamefont
  {Tsoneva}, \citenamefont {Ur}, \citenamefont {Ursescu},\ and\ \citenamefont
  {Zamfir}}]{2020_TanakaKA}%
  \BibitemOpen
  \bibfield  {author} {\bibinfo {author} {\bibfnamefont {K.~A.}\ \bibnamefont
  {Tanaka}}, \bibinfo {author} {\bibfnamefont {K.~M.}\ \bibnamefont {Spohr}},
  \bibinfo {author} {\bibfnamefont {D.~L.}\ \bibnamefont {Balabanski}},
  \bibinfo {author} {\bibfnamefont {S.}~\bibnamefont {Balascuta}}, \bibinfo
  {author} {\bibfnamefont {L.}~\bibnamefont {Capponi}}, \bibinfo {author}
  {\bibfnamefont {M.~O.}\ \bibnamefont {Cernaianu}}, \bibinfo {author}
  {\bibfnamefont {M.}~\bibnamefont {Cuciuc}}, \bibinfo {author} {\bibfnamefont
  {A.}~\bibnamefont {Cucoanes}}, \bibinfo {author} {\bibfnamefont
  {I.}~\bibnamefont {Dancus}}, \bibinfo {author} {\bibfnamefont
  {A.}~\bibnamefont {Dhal}}, \bibinfo {author} {\bibfnamefont {B.}~\bibnamefont
  {Diaconescu}}, \bibinfo {author} {\bibfnamefont {D.}~\bibnamefont {Doria}},
  \bibinfo {author} {\bibfnamefont {P.}~\bibnamefont {Ghenuche}}, \bibinfo
  {author} {\bibfnamefont {D.~G.}\ \bibnamefont {Ghita}}, \bibinfo {author}
  {\bibfnamefont {S.}~\bibnamefont {Kisyov}}, \bibinfo {author} {\bibfnamefont
  {V.}~\bibnamefont {Nastasa}}, \bibinfo {author} {\bibfnamefont {J.~F.}\
  \bibnamefont {Ong}}, \bibinfo {author} {\bibfnamefont {F.}~\bibnamefont
  {Rotaru}}, \bibinfo {author} {\bibfnamefont {D.}~\bibnamefont {Sangwan}},
  \bibinfo {author} {\bibfnamefont {P.-A.}\ \bibnamefont {Söderström}},
  \bibinfo {author} {\bibfnamefont {D.}~\bibnamefont {Stutman}}, \bibinfo
  {author} {\bibfnamefont {G.}~\bibnamefont {Suliman}}, \bibinfo {author}
  {\bibfnamefont {O.}~\bibnamefont {Tesileanu}}, \bibinfo {author}
  {\bibfnamefont {L.}~\bibnamefont {Tudor}}, \bibinfo {author} {\bibfnamefont
  {N.}~\bibnamefont {Tsoneva}}, \bibinfo {author} {\bibfnamefont {C.~A.}\
  \bibnamefont {Ur}}, \bibinfo {author} {\bibfnamefont {D.}~\bibnamefont
  {Ursescu}},\ and\ \bibinfo {author} {\bibfnamefont {N.~V.}\ \bibnamefont
  {Zamfir}},\ }\bibfield  {title} {\bibinfo {title} {Current status and
  highlights of the eli-np research program},\ }\href
  {https://doi.org/10.1063/1.5093535} {\bibfield  {journal} {\bibinfo
  {journal} {Matter Radiat. at Extremes}\ }\textbf {\bibinfo {volume} {5}},\
  \bibinfo {pages} {024402} (\bibinfo {year} {2020})}\BibitemShut {NoStop}%
\bibitem [{\citenamefont {Li}\ \emph {et~al.}(2021)\citenamefont {Li},
  \citenamefont {Kato},\ and\ \citenamefont {Kawanaka}}]{2021_LiZ}%
  \BibitemOpen
  \bibfield  {author} {\bibinfo {author} {\bibfnamefont {Z.}~\bibnamefont
  {Li}}, \bibinfo {author} {\bibfnamefont {Y.}~\bibnamefont {Kato}},\ and\
  \bibinfo {author} {\bibfnamefont {J.}~\bibnamefont {Kawanaka}},\ }\bibfield
  {title} {\bibinfo {title} {Simulating an ultra-broadband concept for
  exawatt-class lasers},\ }\bibfield  {journal} {\bibinfo  {journal} {Sci.
  Rep.}\ }\textbf {\bibinfo {volume} {11}},\ \href
  {https://doi.org/10.1038/s41598-020-80435-6} {10.1038/s41598-020-80435-6}
  (\bibinfo {year} {2021})\BibitemShut {NoStop}%
\bibitem [{\citenamefont {Bulanov}\ \emph {et~al.}(2006)\citenamefont
  {Bulanov}, \citenamefont {Esirkepov}, \citenamefont {Kamenets},\ and\
  \citenamefont {Pegoraro}}]{2006_BulanovSS}%
  \BibitemOpen
  \bibfield  {author} {\bibinfo {author} {\bibfnamefont {S.~S.}\ \bibnamefont
  {Bulanov}}, \bibinfo {author} {\bibfnamefont {T.~Z.}\ \bibnamefont
  {Esirkepov}}, \bibinfo {author} {\bibfnamefont {F.~F.}\ \bibnamefont
  {Kamenets}},\ and\ \bibinfo {author} {\bibfnamefont {F.}~\bibnamefont
  {Pegoraro}},\ }\bibfield  {title} {\bibinfo {title} {Single-cycle
  high-intensity electromagnetic pulse generation in the interaction of a
  plasma wakefield with regular nonlinear structures},\ }\href
  {https://doi.org/10.1103/PhysRevE.73.036408} {\bibfield  {journal} {\bibinfo
  {journal} {Phys. Rev. E}\ }\textbf {\bibinfo {volume} {73}},\ \bibinfo
  {pages} {036408} (\bibinfo {year} {2006})}\BibitemShut {NoStop}%
\bibitem [{\citenamefont {Voronin}\ \emph {et~al.}(2013)\citenamefont
  {Voronin}, \citenamefont {Zheltikov}, \citenamefont {Ditmire}, \citenamefont
  {Rus},\ and\ \citenamefont {Korn}}]{2013_VoroninAA}%
  \BibitemOpen
  \bibfield  {author} {\bibinfo {author} {\bibfnamefont {A.~A.}\ \bibnamefont
  {Voronin}}, \bibinfo {author} {\bibfnamefont {A.~M.}\ \bibnamefont
  {Zheltikov}}, \bibinfo {author} {\bibfnamefont {T.}~\bibnamefont {Ditmire}},
  \bibinfo {author} {\bibfnamefont {B.}~\bibnamefont {Rus}},\ and\ \bibinfo
  {author} {\bibfnamefont {G.}~\bibnamefont {Korn}},\ }\bibfield  {title}
  {\bibinfo {title} {Subexawatt few-cycle lightwave generation via
  multipetawatt pulse compression},\ }\href
  {https://doi.org/https://doi.org/10.1016/j.optcom.2012.10.057} {\bibfield
  {journal} {\bibinfo  {journal} {Opt. Commun.}\ }\textbf {\bibinfo {volume}
  {291}},\ \bibinfo {pages} {299} (\bibinfo {year} {2013})}\BibitemShut
  {NoStop}%
\bibitem [{\citenamefont {Ouillé}\ \emph {et~al.}(2020)\citenamefont
  {Ouillé}, \citenamefont {Vernier}, \citenamefont {Böhle}, \citenamefont
  {Bocoum}, \citenamefont {Jullien}, \citenamefont {Lozano}, \citenamefont
  {Rousseau}, \citenamefont {Cheng}, \citenamefont {Gustas}, \citenamefont
  {Blumenstein}, \citenamefont {Simon}, \citenamefont {Haessler}, \citenamefont
  {Faure}, \citenamefont {Tamas},\ and\ \citenamefont
  {Lopez-Martens}}]{2020_OuilleM}%
  \BibitemOpen
  \bibfield  {author} {\bibinfo {author} {\bibfnamefont {M.}~\bibnamefont
  {Ouillé}}, \bibinfo {author} {\bibfnamefont {A.}~\bibnamefont {Vernier}},
  \bibinfo {author} {\bibfnamefont {F.}~\bibnamefont {Böhle}}, \bibinfo
  {author} {\bibfnamefont {M.}~\bibnamefont {Bocoum}}, \bibinfo {author}
  {\bibfnamefont {A.}~\bibnamefont {Jullien}}, \bibinfo {author} {\bibfnamefont
  {M.}~\bibnamefont {Lozano}}, \bibinfo {author} {\bibfnamefont {J.~P.}\
  \bibnamefont {Rousseau}}, \bibinfo {author} {\bibfnamefont {Z.}~\bibnamefont
  {Cheng}}, \bibinfo {author} {\bibfnamefont {D.}~\bibnamefont {Gustas}},
  \bibinfo {author} {\bibfnamefont {A.}~\bibnamefont {Blumenstein}}, \bibinfo
  {author} {\bibfnamefont {P.}~\bibnamefont {Simon}}, \bibinfo {author}
  {\bibfnamefont {S.}~\bibnamefont {Haessler}}, \bibinfo {author}
  {\bibfnamefont {J.}~\bibnamefont {Faure}}, \bibinfo {author} {\bibfnamefont
  {N.}~\bibnamefont {Tamas}},\ and\ \bibinfo {author} {\bibfnamefont
  {R.}~\bibnamefont {Lopez-Martens}},\ }\bibfield  {title} {\bibinfo {title}
  {Relativistic-intensity near-single-cycle light waveforms at khz repetition
  rate},\ }\bibfield  {journal} {\bibinfo  {journal} {Light Sci. Appl.}\
  }\textbf {\bibinfo {volume} {9}},\ \href
  {https://doi.org/10.1038/s41377-020-0280-5} {10.1038/s41377-020-0280-5}
  (\bibinfo {year} {2020})\BibitemShut {NoStop}%
\bibitem [{\citenamefont {Mourou}\ \emph {et~al.}(2006)\citenamefont {Mourou},
  \citenamefont {Tajima},\ and\ \citenamefont {Bulanov}}]{2006_MourouG}%
  \BibitemOpen
  \bibfield  {author} {\bibinfo {author} {\bibfnamefont {G.~A.}\ \bibnamefont
  {Mourou}}, \bibinfo {author} {\bibfnamefont {T.}~\bibnamefont {Tajima}},\
  and\ \bibinfo {author} {\bibfnamefont {S.~V.}\ \bibnamefont {Bulanov}},\
  }\bibfield  {title} {\bibinfo {title} {Optics in the relativistic regime},\
  }\href {https://doi.org/10.1103/RevModPhys.78.309} {\bibfield  {journal}
  {\bibinfo  {journal} {Rev. Mod. Phys.}\ }\textbf {\bibinfo {volume} {78}},\
  \bibinfo {pages} {309} (\bibinfo {year} {2006})}\BibitemShut {NoStop}%
\bibitem [{\citenamefont {Bulanov}\ \emph {et~al.}(2015)\citenamefont
  {Bulanov}, \citenamefont {Esirkepov}, \citenamefont {Kando}, \citenamefont
  {Koga}, \citenamefont {Kondo},\ and\ \citenamefont {Korn}}]{2015_BulanovSV}%
  \BibitemOpen
  \bibfield  {author} {\bibinfo {author} {\bibfnamefont {S.~V.}\ \bibnamefont
  {Bulanov}}, \bibinfo {author} {\bibfnamefont {T.~Z.}\ \bibnamefont
  {Esirkepov}}, \bibinfo {author} {\bibfnamefont {M.}~\bibnamefont {Kando}},
  \bibinfo {author} {\bibfnamefont {J.}~\bibnamefont {Koga}}, \bibinfo {author}
  {\bibfnamefont {K.}~\bibnamefont {Kondo}},\ and\ \bibinfo {author}
  {\bibfnamefont {G.}~\bibnamefont {Korn}},\ }\bibfield  {title} {\bibinfo
  {title} {On the problems of relativistic laboratory astrophysics and
  fundamental physics with super powerful lasers},\ }\href
  {https://doi.org/10.1134/S1063780X15010018} {\bibfield  {journal} {\bibinfo
  {journal} {Plasma Phys. Rep.}\ }\textbf {\bibinfo {volume} {41}},\ \bibinfo
  {pages} {1} (\bibinfo {year} {2015})}\BibitemShut {NoStop}%
\bibitem [{\citenamefont {Rees}\ and\ \citenamefont
  {Mészáros}(1992)}]{1992_ReesMJ}%
  \BibitemOpen
  \bibfield  {author} {\bibinfo {author} {\bibfnamefont {M.~J.}\ \bibnamefont
  {Rees}}\ and\ \bibinfo {author} {\bibfnamefont {P.}~\bibnamefont
  {Mészáros}},\ }\bibfield  {title} {\bibinfo {title} {Relativistic
  fireballs: energy conversion and time-scales},\ }\href
  {https://doi.org/10.1093/mnras/258.1.41P} {\bibfield  {journal} {\bibinfo
  {journal} {Mon. Not. R. Astron. Soc.}\ }\textbf {\bibinfo {volume} {258}},\
  \bibinfo {pages} {41P} (\bibinfo {year} {1992})}\BibitemShut {NoStop}%
\bibitem [{\citenamefont {Philippov}\ and\ \citenamefont
  {Spitkovsky}(2018)}]{2018_PhilippovAA}%
  \BibitemOpen
  \bibfield  {author} {\bibinfo {author} {\bibfnamefont {A.~A.}\ \bibnamefont
  {Philippov}}\ and\ \bibinfo {author} {\bibfnamefont {A.}~\bibnamefont
  {Spitkovsky}},\ }\bibfield  {title} {\bibinfo {title} {Ab-initio pulsar
  magnetosphere: Particle acceleration in oblique rotators and high-energy
  emission modeling},\ }\href {https://doi.org/10.3847/1538-4357/aaabbc}
  {\bibfield  {journal} {\bibinfo  {journal} {Astrophys. J.}\ }\textbf
  {\bibinfo {volume} {855}},\ \bibinfo {pages} {94} (\bibinfo {year}
  {2018})}\BibitemShut {NoStop}%
\bibitem [{\citenamefont {Eliasson}\ and\ \citenamefont
  {Liu}(2013)}]{2013_EliassonB}%
  \BibitemOpen
  \bibfield  {author} {\bibinfo {author} {\bibfnamefont {B.}~\bibnamefont
  {Eliasson}}\ and\ \bibinfo {author} {\bibfnamefont {C.~S.}\ \bibnamefont
  {Liu}},\ }\bibfield  {title} {\bibinfo {title} {An electromagnetic gamma-ray
  free electron laser},\ }\href {https://doi.org/10.1017/S0022377813000779}
  {\bibfield  {journal} {\bibinfo  {journal} {J. Plasma Phys.}\ }\textbf
  {\bibinfo {volume} {79}},\ \bibinfo {pages} {995–998} (\bibinfo {year}
  {2013})}\BibitemShut {NoStop}%
\bibitem [{\citenamefont {Nedorezov}\ \emph {et~al.}(2004)\citenamefont
  {Nedorezov}, \citenamefont {Turinge},\ and\ \citenamefont
  {Shatunov}}]{2004_NedorezovVG}%
  \BibitemOpen
  \bibfield  {author} {\bibinfo {author} {\bibfnamefont {V.~G.}\ \bibnamefont
  {Nedorezov}}, \bibinfo {author} {\bibfnamefont {A.~A.}\ \bibnamefont
  {Turinge}},\ and\ \bibinfo {author} {\bibfnamefont {Y.~M.}\ \bibnamefont
  {Shatunov}},\ }\bibfield  {title} {\bibinfo {title} {Photonuclear experiments
  with compton-backscattered gamma beams},\ }\href
  {https://doi.org/10.1070/pu2004v047n04abeh001743} {\bibfield  {journal}
  {\bibinfo  {journal} {Phys.-Uspekhi}\ }\textbf {\bibinfo {volume} {47}},\
  \bibinfo {pages} {341} (\bibinfo {year} {2004})}\BibitemShut {NoStop}%
\bibitem [{\citenamefont {Ridgers}\ \emph {et~al.}(2013)\citenamefont
  {Ridgers}, \citenamefont {Brady}, \citenamefont {Duclous}, \citenamefont
  {Kirk}, \citenamefont {Bennett}, \citenamefont {Arber},\ and\ \citenamefont
  {Bell}}]{2013_RidgersCP}%
  \BibitemOpen
  \bibfield  {author} {\bibinfo {author} {\bibfnamefont {C.~P.}\ \bibnamefont
  {Ridgers}}, \bibinfo {author} {\bibfnamefont {C.~S.}\ \bibnamefont {Brady}},
  \bibinfo {author} {\bibfnamefont {R.}~\bibnamefont {Duclous}}, \bibinfo
  {author} {\bibfnamefont {J.~G.}\ \bibnamefont {Kirk}}, \bibinfo {author}
  {\bibfnamefont {K.}~\bibnamefont {Bennett}}, \bibinfo {author} {\bibfnamefont
  {T.~D.}\ \bibnamefont {Arber}},\ and\ \bibinfo {author} {\bibfnamefont
  {A.~R.}\ \bibnamefont {Bell}},\ }\bibfield  {title} {\bibinfo {title} {Dense
  electron-positron plasmas and bursts of gamma-rays from laser-generated
  quantum electrodynamic plasmas},\ }\href {https://doi.org/10.1063/1.4801513}
  {\bibfield  {journal} {\bibinfo  {journal} {Phys. Plasmas}\ }\textbf
  {\bibinfo {volume} {20}},\ \bibinfo {pages} {056701} (\bibinfo {year}
  {2013})}\BibitemShut {NoStop}%
\bibitem [{\citenamefont {Lezhnin}\ \emph {et~al.}(2018)\citenamefont
  {Lezhnin}, \citenamefont {Sasorov}, \citenamefont {Korn},\ and\ \citenamefont
  {Bulanov}}]{2018_LezhninKV}%
  \BibitemOpen
  \bibfield  {author} {\bibinfo {author} {\bibfnamefont {K.~V.}\ \bibnamefont
  {Lezhnin}}, \bibinfo {author} {\bibfnamefont {P.~V.}\ \bibnamefont
  {Sasorov}}, \bibinfo {author} {\bibfnamefont {G.}~\bibnamefont {Korn}},\ and\
  \bibinfo {author} {\bibfnamefont {S.~V.}\ \bibnamefont {Bulanov}},\
  }\bibfield  {title} {\bibinfo {title} {High power gamma flare generation in
  multi-petawatt laser interaction with tailored targets},\ }\href
  {https://doi.org/10.1063/1.5062849} {\bibfield  {journal} {\bibinfo
  {journal} {Phys. Plasmas}\ }\textbf {\bibinfo {volume} {25}},\ \bibinfo
  {pages} {123105} (\bibinfo {year} {2018})}\BibitemShut {NoStop}%
\bibitem [{\citenamefont {Ritus}(1970)}]{1970_RitusVI}%
  \BibitemOpen
  \bibfield  {author} {\bibinfo {author} {\bibfnamefont {V.~I.}\ \bibnamefont
  {Ritus}},\ }\bibfield  {title} {\bibinfo {title} {Radiative effects and their
  enhancement in an intense electromagnetic field},\ }\href
  {http://jetp.ac.ru/cgi-bin/dn/e_030_06_1181.pdf} {\bibfield  {journal}
  {\bibinfo  {journal} {J. Exp. Theor. Phys.}\ }\textbf {\bibinfo {volume}
  {30}},\ \bibinfo {pages} {1181} (\bibinfo {year} {1970})}\BibitemShut
  {NoStop}%
\bibitem [{\citenamefont {Nakamura}\ \emph {et~al.}(2012)\citenamefont
  {Nakamura}, \citenamefont {Koga}, \citenamefont {Esirkepov}, \citenamefont
  {Kando}, \citenamefont {Korn},\ and\ \citenamefont
  {Bulanov}}]{2012_NakamuraT}%
  \BibitemOpen
  \bibfield  {author} {\bibinfo {author} {\bibfnamefont {T.}~\bibnamefont
  {Nakamura}}, \bibinfo {author} {\bibfnamefont {J.~K.}\ \bibnamefont {Koga}},
  \bibinfo {author} {\bibfnamefont {T.~Z.}\ \bibnamefont {Esirkepov}}, \bibinfo
  {author} {\bibfnamefont {M.}~\bibnamefont {Kando}}, \bibinfo {author}
  {\bibfnamefont {G.}~\bibnamefont {Korn}},\ and\ \bibinfo {author}
  {\bibfnamefont {S.~V.}\ \bibnamefont {Bulanov}},\ }\bibfield  {title}
  {\bibinfo {title} {High-power $\ensuremath{\gamma}$-ray flash generation in
  ultraintense laser-plasma interactions},\ }\href
  {https://doi.org/10.1103/PhysRevLett.108.195001} {\bibfield  {journal}
  {\bibinfo  {journal} {Phys. Rev. Lett.}\ }\textbf {\bibinfo {volume} {108}},\
  \bibinfo {pages} {195001} (\bibinfo {year} {2012})}\BibitemShut {NoStop}%
\bibitem [{\citenamefont {Ridgers}\ \emph {et~al.}(2012)\citenamefont
  {Ridgers}, \citenamefont {Brady}, \citenamefont {Duclous}, \citenamefont
  {Kirk}, \citenamefont {Bennett}, \citenamefont {Arber}, \citenamefont
  {Robinson},\ and\ \citenamefont {Bell}}]{2012_RidgersCP}%
  \BibitemOpen
  \bibfield  {author} {\bibinfo {author} {\bibfnamefont {C.~P.}\ \bibnamefont
  {Ridgers}}, \bibinfo {author} {\bibfnamefont {C.~S.}\ \bibnamefont {Brady}},
  \bibinfo {author} {\bibfnamefont {R.}~\bibnamefont {Duclous}}, \bibinfo
  {author} {\bibfnamefont {J.~G.}\ \bibnamefont {Kirk}}, \bibinfo {author}
  {\bibfnamefont {K.}~\bibnamefont {Bennett}}, \bibinfo {author} {\bibfnamefont
  {T.~D.}\ \bibnamefont {Arber}}, \bibinfo {author} {\bibfnamefont {A.~P.~L.}\
  \bibnamefont {Robinson}},\ and\ \bibinfo {author} {\bibfnamefont {A.~R.}\
  \bibnamefont {Bell}},\ }\bibfield  {title} {\bibinfo {title} {Dense
  electron-positron plasmas and ultraintense $\ensuremath{\gamma}$ rays from
  laser-irradiated solids},\ }\href
  {https://doi.org/10.1103/PhysRevLett.108.165006} {\bibfield  {journal}
  {\bibinfo  {journal} {Phys. Rev. Lett.}\ }\textbf {\bibinfo {volume} {108}},\
  \bibinfo {pages} {165006} (\bibinfo {year} {2012})}\BibitemShut {NoStop}%
\bibitem [{\citenamefont {Ehlotzky}\ \emph {et~al.}(2009)\citenamefont
  {Ehlotzky}, \citenamefont {Krajewska},\ and\ \citenamefont
  {Kami{\'{n}}ski}}]{2009_EhlotzkyF}%
  \BibitemOpen
  \bibfield  {author} {\bibinfo {author} {\bibfnamefont {F.}~\bibnamefont
  {Ehlotzky}}, \bibinfo {author} {\bibfnamefont {K.}~\bibnamefont
  {Krajewska}},\ and\ \bibinfo {author} {\bibfnamefont {J.~Z.}\ \bibnamefont
  {Kami{\'{n}}ski}},\ }\bibfield  {title} {\bibinfo {title} {Fundamental
  processes of quantum electrodynamics in laser fields of relativistic power},\
  }\href {https://doi.org/10.1088/0034-4885/72/4/046401} {\bibfield  {journal}
  {\bibinfo  {journal} {Rep. Prog. Phys.}\ }\textbf {\bibinfo {volume} {72}},\
  \bibinfo {pages} {046401} (\bibinfo {year} {2009})}\BibitemShut {NoStop}%
\bibitem [{\citenamefont {Vranic}\ \emph {et~al.}(2016)\citenamefont {Vranic},
  \citenamefont {Grismayer}, \citenamefont {Fonseca},\ and\ \citenamefont
  {Silva}}]{2016_VranicM}%
  \BibitemOpen
  \bibfield  {author} {\bibinfo {author} {\bibfnamefont {M.}~\bibnamefont
  {Vranic}}, \bibinfo {author} {\bibfnamefont {T.}~\bibnamefont {Grismayer}},
  \bibinfo {author} {\bibfnamefont {R.~A.}\ \bibnamefont {Fonseca}},\ and\
  \bibinfo {author} {\bibfnamefont {L.~O.}\ \bibnamefont {Silva}},\ }\bibfield
  {title} {\bibinfo {title} {Electron{\textendash}positron cascades in
  multiple-laser optical traps},\ }\href
  {https://doi.org/10.1088/0741-3335/59/1/014040} {\bibfield  {journal}
  {\bibinfo  {journal} {Plasma Phys. Control. Fusion}\ }\textbf {\bibinfo
  {volume} {59}},\ \bibinfo {pages} {014040} (\bibinfo {year}
  {2016})}\BibitemShut {NoStop}%
\bibitem [{\citenamefont {Gong}\ \emph {et~al.}(2017)\citenamefont {Gong},
  \citenamefont {Hu}, \citenamefont {Shou}, \citenamefont {Qiao}, \citenamefont
  {Chen}, \citenamefont {He}, \citenamefont {Bulanov}, \citenamefont
  {Esirkepov}, \citenamefont {Bulanov},\ and\ \citenamefont
  {Yan}}]{2017_GongZ}%
  \BibitemOpen
  \bibfield  {author} {\bibinfo {author} {\bibfnamefont {Z.}~\bibnamefont
  {Gong}}, \bibinfo {author} {\bibfnamefont {R.~H.}\ \bibnamefont {Hu}},
  \bibinfo {author} {\bibfnamefont {Y.~R.}\ \bibnamefont {Shou}}, \bibinfo
  {author} {\bibfnamefont {B.}~\bibnamefont {Qiao}}, \bibinfo {author}
  {\bibfnamefont {C.~E.}\ \bibnamefont {Chen}}, \bibinfo {author}
  {\bibfnamefont {X.~T.}\ \bibnamefont {He}}, \bibinfo {author} {\bibfnamefont
  {S.~S.}\ \bibnamefont {Bulanov}}, \bibinfo {author} {\bibfnamefont {T.~Z.}\
  \bibnamefont {Esirkepov}}, \bibinfo {author} {\bibfnamefont {S.~V.}\
  \bibnamefont {Bulanov}},\ and\ \bibinfo {author} {\bibfnamefont {X.~Q.}\
  \bibnamefont {Yan}},\ }\bibfield  {title} {\bibinfo {title} {High-efficiency
  $\ensuremath{\gamma}$-ray flash generation via multiple-laser scattering in
  ponderomotive potential well},\ }\href
  {https://doi.org/10.1103/PhysRevE.95.013210} {\bibfield  {journal} {\bibinfo
  {journal} {Phys. Rev. E}\ }\textbf {\bibinfo {volume} {95}},\ \bibinfo
  {pages} {013210} (\bibinfo {year} {2017})}\BibitemShut {NoStop}%
\bibitem [{\citenamefont {Magnusson}\ \emph {et~al.}(2019)\citenamefont
  {Magnusson}, \citenamefont {Gonoskov}, \citenamefont {Marklund},
  \citenamefont {Esirkepov}, \citenamefont {Koga}, \citenamefont {Kondo},
  \citenamefont {Kando}, \citenamefont {Bulanov}, \citenamefont {Korn},\ and\
  \citenamefont {Bulanov}}]{2019_MagnussonJ}%
  \BibitemOpen
  \bibfield  {author} {\bibinfo {author} {\bibfnamefont {J.}~\bibnamefont
  {Magnusson}}, \bibinfo {author} {\bibfnamefont {A.}~\bibnamefont {Gonoskov}},
  \bibinfo {author} {\bibfnamefont {M.}~\bibnamefont {Marklund}}, \bibinfo
  {author} {\bibfnamefont {T.~Z.}\ \bibnamefont {Esirkepov}}, \bibinfo {author}
  {\bibfnamefont {J.~K.}\ \bibnamefont {Koga}}, \bibinfo {author}
  {\bibfnamefont {K.}~\bibnamefont {Kondo}}, \bibinfo {author} {\bibfnamefont
  {M.}~\bibnamefont {Kando}}, \bibinfo {author} {\bibfnamefont {S.~V.}\
  \bibnamefont {Bulanov}}, \bibinfo {author} {\bibfnamefont {G.}~\bibnamefont
  {Korn}},\ and\ \bibinfo {author} {\bibfnamefont {S.~S.}\ \bibnamefont
  {Bulanov}},\ }\bibfield  {title} {\bibinfo {title} {Laser-particle collider
  for multi-gev photon production},\ }\href
  {https://doi.org/10.1103/PhysRevLett.122.254801} {\bibfield  {journal}
  {\bibinfo  {journal} {Phys. Rev. Lett.}\ }\textbf {\bibinfo {volume} {122}},\
  \bibinfo {pages} {254801} (\bibinfo {year} {2019})}\BibitemShut {NoStop}%
\bibitem [{\citenamefont {Sampath}\ \emph {et~al.}(2021)\citenamefont
  {Sampath}, \citenamefont {Davoine}, \citenamefont {Corde}, \citenamefont
  {Gremillet}, \citenamefont {Gilljohann}, \citenamefont {Sangal},
  \citenamefont {Keitel}, \citenamefont {Ariniello}, \citenamefont {Cary},
  \citenamefont {Ekerfelt}, \citenamefont {Emma}, \citenamefont {Fiuza},
  \citenamefont {Fujii}, \citenamefont {Hogan}, \citenamefont {Joshi},
  \citenamefont {Knetsch}, \citenamefont {Kononenko}, \citenamefont {Lee},
  \citenamefont {Litos}, \citenamefont {Marsh}, \citenamefont {Nie},
  \citenamefont {O'Shea}, \citenamefont {Peterson}, \citenamefont {Claveria},
  \citenamefont {Storey}, \citenamefont {Wu}, \citenamefont {Xu}, \citenamefont
  {Zhang},\ and\ \citenamefont {Tamburini}}]{2021_SampathA}%
  \BibitemOpen
  \bibfield  {author} {\bibinfo {author} {\bibfnamefont {A.}~\bibnamefont
  {Sampath}}, \bibinfo {author} {\bibfnamefont {X.}~\bibnamefont {Davoine}},
  \bibinfo {author} {\bibfnamefont {S.}~\bibnamefont {Corde}}, \bibinfo
  {author} {\bibfnamefont {L.}~\bibnamefont {Gremillet}}, \bibinfo {author}
  {\bibfnamefont {M.}~\bibnamefont {Gilljohann}}, \bibinfo {author}
  {\bibfnamefont {M.}~\bibnamefont {Sangal}}, \bibinfo {author} {\bibfnamefont
  {C.~H.}\ \bibnamefont {Keitel}}, \bibinfo {author} {\bibfnamefont
  {R.}~\bibnamefont {Ariniello}}, \bibinfo {author} {\bibfnamefont
  {J.}~\bibnamefont {Cary}}, \bibinfo {author} {\bibfnamefont {H.}~\bibnamefont
  {Ekerfelt}}, \bibinfo {author} {\bibfnamefont {C.}~\bibnamefont {Emma}},
  \bibinfo {author} {\bibfnamefont {F.}~\bibnamefont {Fiuza}}, \bibinfo
  {author} {\bibfnamefont {H.}~\bibnamefont {Fujii}}, \bibinfo {author}
  {\bibfnamefont {M.}~\bibnamefont {Hogan}}, \bibinfo {author} {\bibfnamefont
  {C.}~\bibnamefont {Joshi}}, \bibinfo {author} {\bibfnamefont
  {A.}~\bibnamefont {Knetsch}}, \bibinfo {author} {\bibfnamefont
  {O.}~\bibnamefont {Kononenko}}, \bibinfo {author} {\bibfnamefont
  {V.}~\bibnamefont {Lee}}, \bibinfo {author} {\bibfnamefont {M.}~\bibnamefont
  {Litos}}, \bibinfo {author} {\bibfnamefont {K.}~\bibnamefont {Marsh}},
  \bibinfo {author} {\bibfnamefont {Z.}~\bibnamefont {Nie}}, \bibinfo {author}
  {\bibfnamefont {B.}~\bibnamefont {O'Shea}}, \bibinfo {author} {\bibfnamefont
  {J.~R.}\ \bibnamefont {Peterson}}, \bibinfo {author} {\bibfnamefont
  {P.~S.~M.}\ \bibnamefont {Claveria}}, \bibinfo {author} {\bibfnamefont
  {D.}~\bibnamefont {Storey}}, \bibinfo {author} {\bibfnamefont
  {Y.}~\bibnamefont {Wu}}, \bibinfo {author} {\bibfnamefont {X.}~\bibnamefont
  {Xu}}, \bibinfo {author} {\bibfnamefont {C.}~\bibnamefont {Zhang}},\ and\
  \bibinfo {author} {\bibfnamefont {M.}~\bibnamefont {Tamburini}},\ }\bibfield
  {title} {\bibinfo {title} {Extremely dense gamma-ray pulses in electron
  beam-multifoil collisions},\ }\href
  {https://doi.org/10.1103/PhysRevLett.126.064801} {\bibfield  {journal}
  {\bibinfo  {journal} {Phys. Rev. Lett.}\ }\textbf {\bibinfo {volume} {126}},\
  \bibinfo {pages} {064801} (\bibinfo {year} {2021})}\BibitemShut {NoStop}%
\bibitem [{\citenamefont {Jeong}\ \emph {et~al.}(2015)\citenamefont {Jeong},
  \citenamefont {Weber}, \citenamefont {Le~Garrec}, \citenamefont {Margarone},
  \citenamefont {Mocek},\ and\ \citenamefont {Korn}}]{2015_JeongTM}%
  \BibitemOpen
  \bibfield  {author} {\bibinfo {author} {\bibfnamefont {T.~M.}\ \bibnamefont
  {Jeong}}, \bibinfo {author} {\bibfnamefont {S.}~\bibnamefont {Weber}},
  \bibinfo {author} {\bibfnamefont {B.}~\bibnamefont {Le~Garrec}}, \bibinfo
  {author} {\bibfnamefont {D.}~\bibnamefont {Margarone}}, \bibinfo {author}
  {\bibfnamefont {T.}~\bibnamefont {Mocek}},\ and\ \bibinfo {author}
  {\bibfnamefont {G.}~\bibnamefont {Korn}},\ }\bibfield  {title} {\bibinfo
  {title} {Spatio-temporal modification of femtosecond focal spot under tight
  focusing condition},\ }\href {https://doi.org/10.1364/OE.23.011641}
  {\bibfield  {journal} {\bibinfo  {journal} {Opt. Express}\ }\textbf {\bibinfo
  {volume} {23}},\ \bibinfo {pages} {11641} (\bibinfo {year}
  {2015})}\BibitemShut {NoStop}%
\bibitem [{\citenamefont {Jeong}\ \emph {et~al.}(2018)\citenamefont {Jeong},
  \citenamefont {Bulanov}, \citenamefont {Weber},\ and\ \citenamefont
  {Korn}}]{2018_JeongTM}%
  \BibitemOpen
  \bibfield  {author} {\bibinfo {author} {\bibfnamefont {T.~M.}\ \bibnamefont
  {Jeong}}, \bibinfo {author} {\bibfnamefont {S.~V.}\ \bibnamefont {Bulanov}},
  \bibinfo {author} {\bibfnamefont {S.}~\bibnamefont {Weber}},\ and\ \bibinfo
  {author} {\bibfnamefont {G.}~\bibnamefont {Korn}},\ }\bibfield  {title}
  {\bibinfo {title} {Analysis on the longitudinal field strength formed by
  tightly-focused radially-polarized femtosecond petawatt laser pulse},\ }\href
  {https://doi.org/10.1364/OE.26.033091} {\bibfield  {journal} {\bibinfo
  {journal} {Opt. Express}\ }\textbf {\bibinfo {volume} {26}},\ \bibinfo
  {pages} {33091} (\bibinfo {year} {2018})}\BibitemShut {NoStop}%
\bibitem [{\citenamefont {Mourou}\ \emph {et~al.}(2002)\citenamefont {Mourou},
  \citenamefont {Chang}, \citenamefont {Maksimchuk}, \citenamefont {Nees},
  \citenamefont {Bulanov}, \citenamefont {Bychenkov}, \citenamefont
  {Esirkepov}, \citenamefont {Naumova}, \citenamefont {Pegoraro},\ and\
  \citenamefont {Ruhl}}]{2002_MourouG}%
  \BibitemOpen
  \bibfield  {author} {\bibinfo {author} {\bibfnamefont {G.}~\bibnamefont
  {Mourou}}, \bibinfo {author} {\bibfnamefont {Z.}~\bibnamefont {Chang}},
  \bibinfo {author} {\bibfnamefont {A.}~\bibnamefont {Maksimchuk}}, \bibinfo
  {author} {\bibfnamefont {J.}~\bibnamefont {Nees}}, \bibinfo {author}
  {\bibfnamefont {S.~V.}\ \bibnamefont {Bulanov}}, \bibinfo {author}
  {\bibfnamefont {V.~Y.}\ \bibnamefont {Bychenkov}}, \bibinfo {author}
  {\bibfnamefont {T.~Z.}\ \bibnamefont {Esirkepov}}, \bibinfo {author}
  {\bibfnamefont {N.~M.}\ \bibnamefont {Naumova}}, \bibinfo {author}
  {\bibfnamefont {F.}~\bibnamefont {Pegoraro}},\ and\ \bibinfo {author}
  {\bibfnamefont {H.}~\bibnamefont {Ruhl}},\ }\bibfield  {title} {\bibinfo
  {title} {On the design of experiments for the study of relativistic nonlinear
  optics in the limit of single-cycle pulse duration and single-wavelength spot
  size},\ }\href {https://doi.org/10.1134/1.1434292} {\bibfield  {journal}
  {\bibinfo  {journal} {Plasma Phys. Rep.}\ }\textbf {\bibinfo {volume} {28}},\
  \bibinfo {pages} {12} (\bibinfo {year} {2002})}\BibitemShut {NoStop}%
\bibitem [{\citenamefont {Naumova}\ \emph {et~al.}(2004)\citenamefont
  {Naumova}, \citenamefont {Nees}, \citenamefont {Sokolov}, \citenamefont
  {Hou},\ and\ \citenamefont {Mourou}}]{2004_NaumovaNM}%
  \BibitemOpen
  \bibfield  {author} {\bibinfo {author} {\bibfnamefont {N.~M.}\ \bibnamefont
  {Naumova}}, \bibinfo {author} {\bibfnamefont {J.~A.}\ \bibnamefont {Nees}},
  \bibinfo {author} {\bibfnamefont {I.~V.}\ \bibnamefont {Sokolov}}, \bibinfo
  {author} {\bibfnamefont {B.}~\bibnamefont {Hou}},\ and\ \bibinfo {author}
  {\bibfnamefont {G.~A.}\ \bibnamefont {Mourou}},\ }\bibfield  {title}
  {\bibinfo {title} {Relativistic generation of isolated attosecond pulses in a
  ${\ensuremath{\lambda}}^{3}$ focal volume},\ }\href
  {https://doi.org/10.1103/PhysRevLett.92.063902} {\bibfield  {journal}
  {\bibinfo  {journal} {Phys. Rev. Lett.}\ }\textbf {\bibinfo {volume} {92}},\
  \bibinfo {pages} {063902} (\bibinfo {year} {2004})}\BibitemShut {NoStop}%
\bibitem [{\citenamefont {Arber}\ \emph {et~al.}(2015)\citenamefont {Arber},
  \citenamefont {Bennett}, \citenamefont {Brady}, \citenamefont
  {Lawrence-Douglas}, \citenamefont {Ramsay}, \citenamefont {Sircombe},
  \citenamefont {Gillies}, \citenamefont {Evans}, \citenamefont {Schmitz},
  \citenamefont {Bell},\ and\ \citenamefont {Ridgers}}]{2015_ArberTD}%
  \BibitemOpen
  \bibfield  {author} {\bibinfo {author} {\bibfnamefont {T.~D.}\ \bibnamefont
  {Arber}}, \bibinfo {author} {\bibfnamefont {K.}~\bibnamefont {Bennett}},
  \bibinfo {author} {\bibfnamefont {C.~S.}\ \bibnamefont {Brady}}, \bibinfo
  {author} {\bibfnamefont {A.}~\bibnamefont {Lawrence-Douglas}}, \bibinfo
  {author} {\bibfnamefont {M.~G.}\ \bibnamefont {Ramsay}}, \bibinfo {author}
  {\bibfnamefont {N.~J.}\ \bibnamefont {Sircombe}}, \bibinfo {author}
  {\bibfnamefont {P.}~\bibnamefont {Gillies}}, \bibinfo {author} {\bibfnamefont
  {R.~G.}\ \bibnamefont {Evans}}, \bibinfo {author} {\bibfnamefont
  {H.}~\bibnamefont {Schmitz}}, \bibinfo {author} {\bibfnamefont {A.~R.}\
  \bibnamefont {Bell}},\ and\ \bibinfo {author} {\bibfnamefont {C.~P.}\
  \bibnamefont {Ridgers}},\ }\bibfield  {title} {\bibinfo {title} {Contemporary
  particle-in-cell approach to laser-plasma modelling},\ }\href
  {https://doi.org/10.1088/0741-3335/57/11/113001} {\bibfield  {journal}
  {\bibinfo  {journal} {Plasma Phys. Control. Fusion}\ }\textbf {\bibinfo
  {volume} {57}},\ \bibinfo {pages} {113001} (\bibinfo {year}
  {2015})}\BibitemShut {NoStop}%
\bibitem [{\citenamefont {Ridgers}\ \emph {et~al.}(2014)\citenamefont
  {Ridgers}, \citenamefont {Kirk}, \citenamefont {Duclous}, \citenamefont
  {Blackburn}, \citenamefont {Brady}, \citenamefont {Bennett}, \citenamefont
  {Arber},\ and\ \citenamefont {Bell}}]{2014_RidgersCP}%
  \BibitemOpen
  \bibfield  {author} {\bibinfo {author} {\bibfnamefont {C.~P.}\ \bibnamefont
  {Ridgers}}, \bibinfo {author} {\bibfnamefont {J.~G.}\ \bibnamefont {Kirk}},
  \bibinfo {author} {\bibfnamefont {R.}~\bibnamefont {Duclous}}, \bibinfo
  {author} {\bibfnamefont {T.~G.}\ \bibnamefont {Blackburn}}, \bibinfo {author}
  {\bibfnamefont {C.~S.}\ \bibnamefont {Brady}}, \bibinfo {author}
  {\bibfnamefont {K.}~\bibnamefont {Bennett}}, \bibinfo {author} {\bibfnamefont
  {T.~D.}\ \bibnamefont {Arber}},\ and\ \bibinfo {author} {\bibfnamefont
  {A.~R.}\ \bibnamefont {Bell}},\ }\bibfield  {title} {\bibinfo {title}
  {Modelling gamma-ray photon emission and pair production in high-intensity
  laser–matter interactions},\ }\href
  {https://doi.org/https://doi.org/10.1016/j.jcp.2013.12.007} {\bibfield
  {journal} {\bibinfo  {journal} {J. Comput. Phys}\ }\textbf {\bibinfo {volume}
  {260}},\ \bibinfo {pages} {273} (\bibinfo {year} {2014})}\BibitemShut
  {NoStop}%
\bibitem [{\citenamefont {Higuera}\ and\ \citenamefont
  {Cary}(2017)}]{2017_HigueraAV}%
  \BibitemOpen
  \bibfield  {author} {\bibinfo {author} {\bibfnamefont {A.~V.}\ \bibnamefont
  {Higuera}}\ and\ \bibinfo {author} {\bibfnamefont {J.~R.}\ \bibnamefont
  {Cary}},\ }\bibfield  {title} {\bibinfo {title} {Structure-preserving
  second-order integration of relativistic charged particle trajectories in
  electromagnetic fields},\ }\href {https://doi.org/10.1063/1.4979989}
  {\bibfield  {journal} {\bibinfo  {journal} {Phys. Plasmas}\ }\textbf
  {\bibinfo {volume} {24}},\ \bibinfo {pages} {052104} (\bibinfo {year}
  {2017})}\BibitemShut {NoStop}%
\bibitem [{\citenamefont {Vshivkov}\ \emph {et~al.}(1998)\citenamefont
  {Vshivkov}, \citenamefont {Naumova}, \citenamefont {Pegoraro},\ and\
  \citenamefont {Bulanov}}]{1998_VshivkovVA}%
  \BibitemOpen
  \bibfield  {author} {\bibinfo {author} {\bibfnamefont {V.~A.}\ \bibnamefont
  {Vshivkov}}, \bibinfo {author} {\bibfnamefont {N.~M.}\ \bibnamefont
  {Naumova}}, \bibinfo {author} {\bibfnamefont {F.}~\bibnamefont {Pegoraro}},\
  and\ \bibinfo {author} {\bibfnamefont {S.~V.}\ \bibnamefont {Bulanov}},\
  }\bibfield  {title} {\bibinfo {title} {Nonlinear electrodynamics of the
  interaction of ultra-intense laser pulses with a thin foil},\ }\href
  {https://doi.org/10.1063/1.872961} {\bibfield  {journal} {\bibinfo  {journal}
  {Phys. Plasmas}\ }\textbf {\bibinfo {volume} {5}},\ \bibinfo {pages} {2727}
  (\bibinfo {year} {1998})}\BibitemShut {NoStop}%
\bibitem [{\citenamefont {Bell}\ and\ \citenamefont
  {Kirk}(2008)}]{2008_BellAR}%
  \BibitemOpen
  \bibfield  {author} {\bibinfo {author} {\bibfnamefont {A.~R.}\ \bibnamefont
  {Bell}}\ and\ \bibinfo {author} {\bibfnamefont {J.~G.}\ \bibnamefont
  {Kirk}},\ }\bibfield  {title} {\bibinfo {title} {Possibility of prolific pair
  production with high-power lasers},\ }\href
  {https://doi.org/10.1103/PhysRevLett.101.200403} {\bibfield  {journal}
  {\bibinfo  {journal} {Phys. Rev. Lett.}\ }\textbf {\bibinfo {volume} {101}},\
  \bibinfo {pages} {200403} (\bibinfo {year} {2008})}\BibitemShut {NoStop}%
\end{thebibliography}%

\end{document}